\documentclass[journal]{IEEEtran}

\usepackage{cite}
\usepackage{dcolumn}
\usepackage{multirow}
\usepackage{placeins}

\usepackage[pdftex]{graphicx}
\DeclareGraphicsExtensions{.pdf,.jpeg,.png}

\usepackage{amsmath}

\begin{document}

{\onecolumn
\copyright  2017 IEEE. Personal use of this material is permitted. Permission from IEEE must be obtained for all other
uses, including reprinting/republishing this material for advertising or promotional purposes, collecting new
collected works for resale or redistribution to servers or lists, or reuse of any copyrighted component of this
work in other works.

\medskip
A.~Pikalev, V.~Kobylin and A.~Semenov, "Spectral Study of Glow Discharge With Dusty Structures", 
in \textit{IEEE Transactions on Plasma Science}, vol. 46, no. 4, pp. 698--700, April 2018. \\
doi: 10.1109/TPS.2017.2763742 \\
URL: http://ieeexplore.ieee.org/document/8103344/
}

\twocolumn
\title{Spectral Study of Glow Discharge with Dusty Structures}

\author{Aleksandr~Pikalev,
        Vladimir~Kobylin,
        and~Aleksandr~Semenov
\thanks{The authors are with the Institute of Physics\&Technology,
Petrozavodsk State University, Petrozavodsk, 185910, Russian Federation
(email: pikalev@dims.karelia.ru)}}%

\markboth{IEEE TRANSACTIONS ON PLASMA SCIENCE}%
{Pikalev \MakeLowercase{\textit{et al.}}: Spectral Study of Glow Discharge with Dusty Structures}

\maketitle

\begin{abstract}
We present the results of the experimental study of the influence of dusty structures on plasma. 
We have studied glow discharges in neon and argon with $\text{Al}_2 \text{O}_3$ polydisperse particles 
and melamine-formaldehyde monodisperse particles with the diameter of $\text{4.83} \; \mu m$.
We used optical emission spectroscopy and laser-induced fluorescence methods 
to measure shift of excited atoms populations, which led to similar results. 
In most cases, the population change was negligible as compared to the inaccuracy. 
$\text{Al}_2 \text{O}_3$ particles in argon formed large structures which caused dramatic redistribution of radiation.
In this case, spectral lines and fluorescence became 9--50\% weaker at the tube center after dusty structure formation. 
It might be accounted for by a decrease of electron density due to recombination on the surfaces of the particles. 
Near the tube wall, the radiation became brighter due to an electron temperature increase.
Photographs of the discharge allowed us to calculate the radial distribution of the population shift. 
\end{abstract}


\section{Introduction}

\IEEEPARstart{P}{lasma} with levitated macroparticles is called dusty or complex plasma.
Dusty structures influence plasma conditions.
In a glow discharge, the dusty particles have negative charge because electrons have much higher temperature and 
much lower mass than ions \cite{Fortov2009}. 
Thus the electron density becomes less than the ion density.
Electrons and ions recombine on particle surfaces and the electron temperature increases to compensate these losses.
Metastable atoms also die on the surfaces. 

In papers \cite{Do2008, Do2009, Mitic2009, Layden2011, Alexandrov2013, Stefanovic2017}, 
the influence of dusty particles on excited atoms density 
has been investigated in Ne and Ar RF discharge.
The effect value and even its sign depend on the gas, particle size, and discharge conditions.

In papers \cite{Khakhaev2003, Bulba2006}, spectra were measured for a variety of the dusty structure sizes
in neon and argon glow discharge.
The experiments revealed that the electric field strength in a discharge increased 
at low currents in neon and argon upon the injection of $\text{Al}_2 \text{O}_3$ and zinc particles, whereas
line intensities remained unchanged against the background of experimental errors. 
In contrast, at a current of 2~mA, the effect of particles on the discharge supply voltage was undetectable, 
whereas the intensity of lines increased. 
At all currents in neon, the line intensity increased proportionally to the volume of the structure,
whereas, in contrast, at all currents in argon, the line intensity insignificantly decreased with the structure growth.

In \cite{Kobylin2006}, the influence of the structures on Ne levels population was investigated with laser-induced 
fluorescence in a glow discharge. Dusty particles increased the population, 
but only in one condition set the increasing was statistically significant.
Theoretical analysis of the dust-plasma interaction in a glow discharge was developed in 
\cite{Fedoseev2011,Shumova2014,Shumova2017}.

A detailed review of dusty structures influence on glow discharge plasma can be found in \cite{Polyakov2017}.

\section{Experimental setup}

\begin{figure}[t]
\centering
\includegraphics[width=1\linewidth]{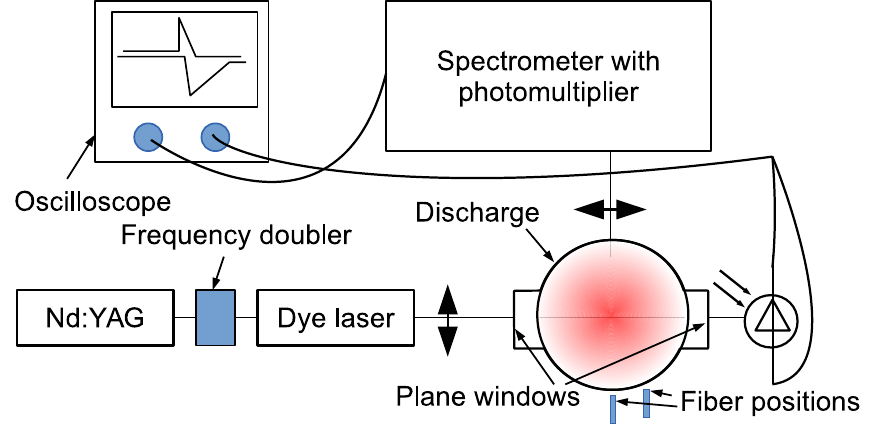}
\caption{The experimental setup}
\label{scheme}
\end{figure}

We used laser-induced fluorescence method and optical emission spectroscopy.
The scheme of the experimental setup is presented in Fig. \ref{scheme}.
The laser-induced fluorescence was used to measure the population of the metastable and resonance energy levels,
which do not have quantum transitions in the visible spectral range.
The method involved 
plasma irradiation with a dye laser turned in resonance with absorption line of an excited atom. 
The atom transited to a more excited state and could relax in several spectral lines.
Hence, laser shot led to fluorescence impulse, and the higher the atom density in the initial excited state was,
the stronger fluorescence pulse was registered.
Our dye laser produced the pulses with the duration of about 10~ns and frequency of 6~Hz. 
The laser line width was of the order of 0.01~nm which was much wider than a spectral line width in the discharge.

The discharge tube (\O ~30~mm) had plane windows that did not distort the laser beam.
The laser beam was positioned near the down border of the windows to pass through the dusty structure. 
Also, the tube contained a stricture (narrowing \O~5~mm) that stabilized the plasma stratification
and created a perturbation for the dusty structure trapping (see Fig.~\ref{photos}C).

We used Avantes Avaspec 2048FT-2-DT spectrometer for emission spectroscopy.
The spectrometer had a fiber input.
We conducted 2 series of experiments.
Firstly, we pointed the fiber to the tube center and, secondly, to an area near the tube wall 
(about 8~mm from the tube center).
The fiber was situated at the same height as the laser beam for laser induced fluorescence.

To measure the emission distribution, we took photos of the discharge.

\section{Results}

\begin{table*}
	\centering
	\caption{Relative Difference between Fluorescence With and Without Particles} \label{width_dif}  
	\begin{tabular}[c]{|r|r|r|r|r|}
		\hline
      	\multirow{2}{60pt}{Gas and particles} & \multirow{2}{100pt}{Wave length of pumping and fluorescence, nm} & \multirow{2}{40pt}{Pressure, Pa} & \multirow{2}{45pt}{Current, mA}& \multirow{2}{50pt}{Difference, \%} \\
      	& &  &	&\\ \hline
		\multirow{4}*{Ne + $\text{Al}_2\text O_3$ (polydisperse)} &\multirow{3}*{693; 614} &\multirow{2}*{100} & 1.5 & $ 0.1 \pm 0.6$\\ \cline{4-5}
		                                                          &                        &                   & 2.0 & $-0.3 \pm 0.5$\\ \cline{3-5}
		                                                          &                        &  150              & 1.5 & $-1.0  \pm 2.9 $\\ \cline{2-5}
		                                                          & 703; 724               &  100              & 1.5 & $ 0.2  \pm 1.0 $\\ \hline
		Ar + $\text{Al}_2 \text O_3$ (polydisperse)               & 696; 727               &   30              & 1.5 & $-9.1  \pm 3.2 $\\ \hline
		\multirow{2}*{Ar + MF \O~$4.83 \mu m$}                    &\multirow{2}*{ 696; 727}&   15              & 3.0 & $ 1.9  \pm 5.5 $\\ \cline{3-5}
		                                                          &                        &   20              & 3.0 & $ 3.7  \pm 3.4 $\\ \hline
	\end{tabular}
\end{table*}

The results of the laser-induced fluorescence have been previously presented at a conference \cite{Kobylin2016}.

We studied glow discharges in neon and argon with 
$\text{Al}_2 \text O_3$ polydisperse particles and melamine-formaldehyde monodisperse particles 
with the diameter of 4.83~$\mu m$.
In neon we investigated levels $2s^2 2p^5 (^2 P^o_{1/2})3s \;\; ^2[1/2]^o \;\; J=1$ (pumping --- 692.9~nm, 
fluorescence --- 614.3~nm) and $2s^2 2p^5(^2P^o_{3/2})3s \;\; ^2[3/2]^o \;\; J=2$ (pumping --- 703.2~nm, 
fluorescence --- 724.5~nm). In argon we dealt with $3s^2 3p^5(^2P^o_{3/2})4s \;\; ^2[3/2]^o \;\; J=2$ level 
(pumping --- 696.5~nm, fluorescence --- 727.3~nm).

The results are presented in table~\ref{width_dif}.
In most cases the population change was negligible as compared to the inaccuracy. 
$\text{Al}_2 \text O_3$ particles caused significant decrease of the fluorescence.

\begin{figure}[t]
\centering
\includegraphics[width=1\linewidth]{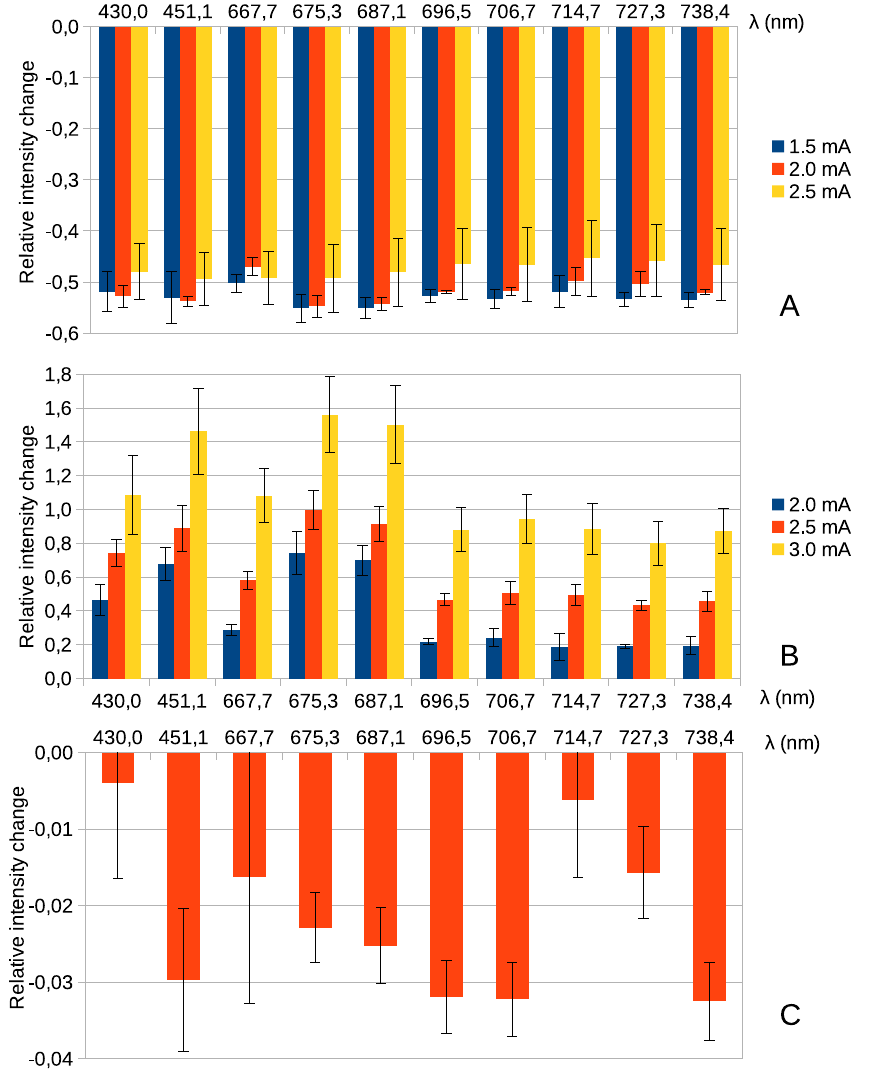}
\caption{Relative intensity change of Ar spectral lines. 
A --- with polydisperse $\text{Al}_2 \text{O}_3$ particles. P~=~30~Pa. The fiber was pointed to the tube center.
B --- with polydisperse $\text{Al}_2 \text{O}_3$ particles. P~=~30~Pa. The fiber was pointed to an area near the tube wall.
C --- with  melamine-formaldehyde particles \O~$4.83 \mu m$ . P~=~15~Pa, I~=~3~mA. The fiber was pointed to the tube center.
}
\label{spectr}
\end{figure}

The spectrum measurements showed the same results. 
In Fig. \ref{spectr}, the relative influence of dusty structures is presented,
which is the value of $\dfrac{I_{dusty} - I_{free}}{I_{free}}$,
where $I_{dusty}$ and $I_{free}$ are spectral line intensities with and without particles, respectively.
The positive values mean that intensity with dusty structure is higher than that without the particles
and negative values mean that the dusty structure decreases the intensity.
The dusty structures cause an intensity decrease  at the tube center and an increase near the wall.

\begin{figure}[t]
\centering
\includegraphics[width=1\linewidth]{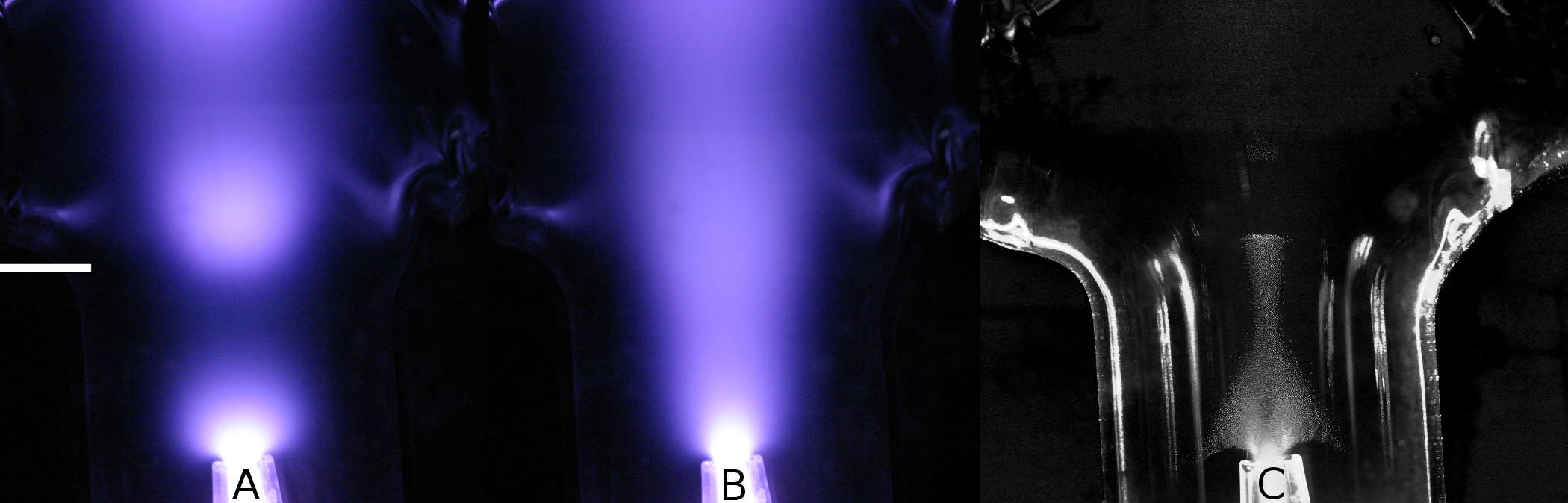}
\caption{\it \small Photographs of the discharge. Ar P~=~30~Pa, I~=~2~mA. 
A --- plasma without particles; 
B --- plasma with $\text{Al}_2 \text O_3$ polydisperse particles;
C --- the dusty structure illuminated with a laser knife.
The plane windows are visible at the top and the narrowing is at the bottom. 
The top of the structure is shadowed with the window border.
A white mark in the left shows the height at which radial distribution was calculated.
}
\label{photos}
\end{figure}

$\text{Al}_2\text O_3$ particles form one elongated structure with a complex shape.
Melamine-formaldehyde particles in our conditions form two shot structures: 
one is just above the narrowing and another structure is at the windows level.
We investigated only the upper ones those are about 8~mm in length and 5~mm in diameter.
Photographs in Fig. \ref{photos} show, that $\text{Al}_2\text O_3$ particles dramatically change the Ar plasma emittance.
In other cases (Ne or Ar with melamine-formaldehyde particles) the effect is not eye-visible.

We have calculated radial radiation distribution with the Abel transform:

\begin{equation}
\varepsilon (r) = - \dfrac{1}{\pi} \int_r^R \dfrac{db/dx}{\sqrt{x^2 - r^2}} dx
\end{equation}

Here $\varepsilon (r)$ is the radiation of elementary volume of the plasma,
$R$ --- the tube radius,
$b(x)$ --- the plasma brightness, and
$x$ is the pixel horizontal coordinate calculated from the discharge image center.

\begin{figure}[t]
\centering
\includegraphics[width=1\linewidth]{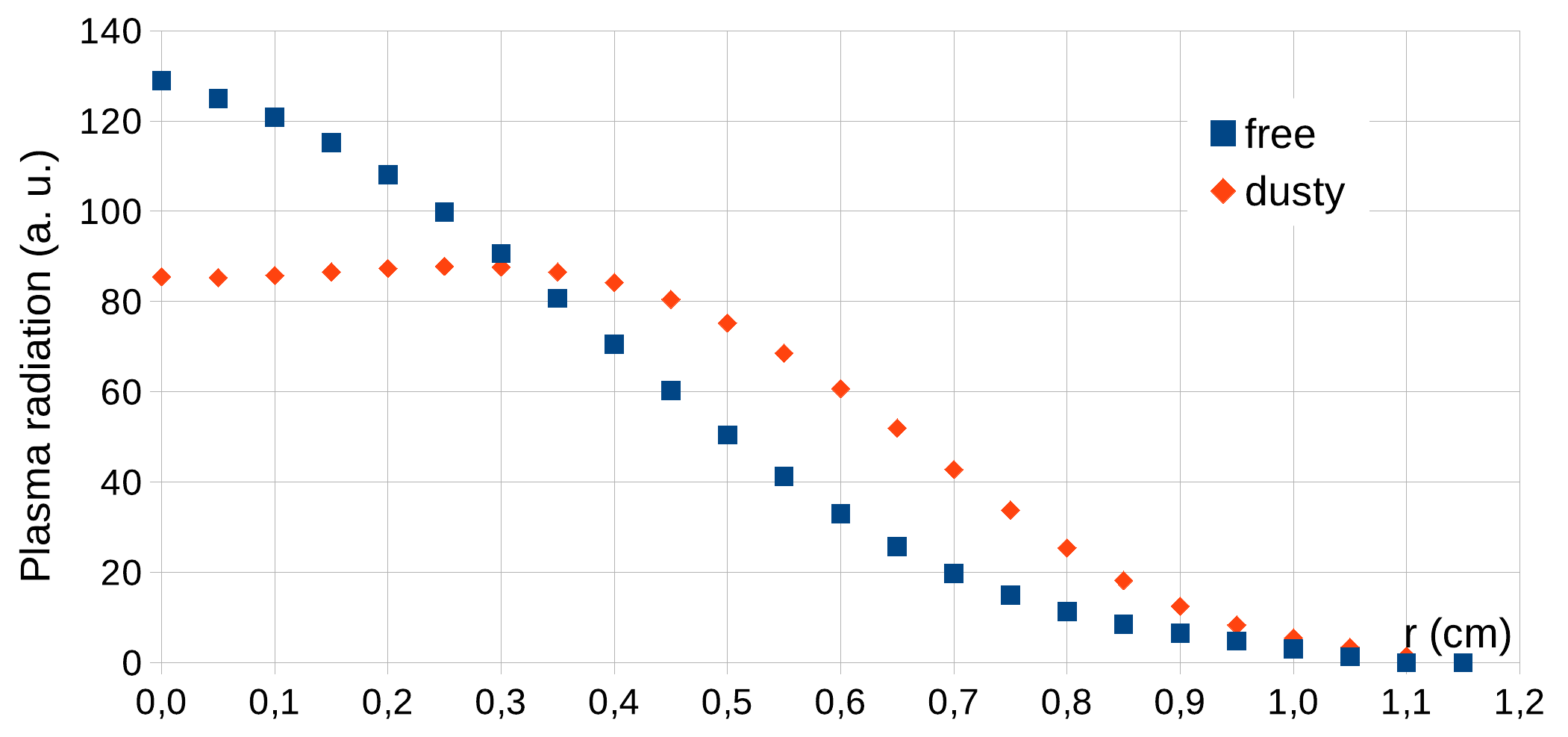}
\caption{Radial distribution of the plasma radiation from Fig.~\ref{photos} in arbitrary units. 
The structure radius at this height is 0.18~cm.}
\label{radial}
\end{figure}

We calculated the distribution (Fig. \ref{radial}) in arbitrary units, because we used pixel values of the photographs as the brightness.

The distribution was calculated at the height slightly lower than laser-induced fluorescence, and 
the spectrum measurements were conducted (see a mark at Fig. \ref{photos}),
because the Abel transform is suitable only for cylindrical symmetrical objects.

\section{Conclusion}
The experiments have shown that dusty structure can greatly change the plasma conditions in positive column,
but the effect depends on a type of gas, discharge conditions, size and concentration of particles and size of the structure.
The structure can completely change the radiation distribution, both radial and axial one.
The radial distribution of plasma radiation with dust particles in Fig. \ref{radial} 
correlates with the distribution of electrons obtained in \cite{Shumova2014} for neon.
It can be explained with electron and ion recombination on the surfaces of the dust,
which leads to an electron density decrease near the tube center, and thus, to a decrease in the excited atoms population.
The electric field and the electron temperature grow to compensate the losses \cite{Polyakov2017}.
That is why the population of excited atoms near the tube wall is greater in dusty plasma than that in plasma without particles.
For the quantitative explanation of the result, 
the strata theory of positive column with a nonuniform radius should be taken into account.

\section*{Acknowledgment}

The work was supported by the Russian Foundation for Basic Research, project no. 16-32-00229.

We appreciate A.~I.~Scherbina for his assistance with the experimental setup, 
and A.~L.~Pergament and V.~I.~Sysun for valuable discussions.
\FloatBarrier

\begin{IEEEbiographynophoto}{Aleksandr Pikalev}
received the B.S. and M.S. degree in information technology from Petrozavodsk State University in 2010 and 2012 respectively.
Today he is finishing PhD thesis in physical electronics.
He is an engineer and a lecturer at the Petrozavodsk State University, Karelia Republic, Russia.
His research interests include dusty plasma physics, spectroscopy and experiment automation.
\end{IEEEbiographynophoto}

\begin{IEEEbiographynophoto}{Vladimir Kobylin}
is a graduate of Petrozavodsk State University of 1982, Department of Physics and Mathematics, specialty --- physics.
At present he is an employee of Petrozavodsk State University, Karelia Republic, Russian Federation.
His current research interests include basic plasma and laser physics and dusty plasmas.
\end{IEEEbiographynophoto}

\begin{IEEEbiographynophoto}{Aleksandr Semenov}
received the Ph.D. degree in physics from the Petrozavodsk State University, Karelia Republic, Russia. 
He is currently an lecturer and engineer at the Petrozavodsk State University.
His research interests include dusty plasma physics, plasma-medicine, biophysics.
\end{IEEEbiographynophoto}

\vfill

\end{document}